\documentclass[apj]{emulateapj}
\journalinfo{\textsc{The Astrophysical Journal}}
\slugcomment{Received 2016 February 28; Accepted 2016 July 12}
\usepackage{url}

\usepackage{apjfonts}
\usepackage{natbib}
\usepackage{booktabs}
\usepackage{epstopdf}
\usepackage{graphicx}

\usepackage[colorlinks,
        hypertexnames=false,
        pagebackref=true,
        linkcolor=blue,    
    citecolor=blue,        
    filecolor=magenta,     
    urlcolor=blue          
]{hyperref}                
\usepackage{amsmath}

\shorttitle{Sunsport rotation as a driver of major solar eruptions}
\shortauthors{Vemareddy et al}
\begin{document}
\title{Sunspot Rotation as a Driver of Major Solar Eruptions in NOAA Active Region 12158}
\author{P.~Vemareddy$^1$, X.~Cheng$^{2}$, and B.~Ravindra$^1$}
\affil{$^1$Indian Institute of Astrophysics, Koramangala, Bangalore-560034, India}
\affil{$^2$School of Astronomy and Space Science, Nanjing University, Nanjing-210023, China} 
\email{vemareddy@iiap.res.in}
\begin{abstract}
We studied the developing conditions of sigmoid structure under the influence of magnetic non-potential characteristics of a rotating sunspot in the active region (AR) 12158. Vector magnetic field measurements from Helioseismic Magnetic Imager and coronal EUV observations from Atmospheric Imaging Assembly reveal that the erupting inverse-S sigmoid had roots in the location of the rotating sunspot. Sunspot rotates at a rate of 0-5deg/h with increasing trend in the first half followed by a decrease. Time evolution of many non-potential parameters had a well correspondence with the sunspot rotation. The evolution of the AR magnetic structure is approximated by a time series of force free equilibria. The NLFFF magnetic structure around the sunspot manifests the observed sigmoid structure. Field lines from the sunspot periphery constitute the body of the sigmoid and those from interior overly the sigmoid similar to a fluxrope structure. While the sunspot is being rotating, two major CME eruptions occurred in the AR. During the first (second) event, the coronal current concentrations enhanced (degraded) consistent with the photospheric net vertical current, however the magnetic energy is released during both the cases. The analysis results suggest that the magnetic connections of the sigmoid are driven by slow motion of sunspot rotation, which transforms to a highly twisted flux rope structure in a dynamical scenario. An exceeding critical twist in the flux rope probably leads to the loss of equilibrium and thus triggering the onset of two eruptions.
\end{abstract}

\section{Introduction}
\label{Intro}
It is generally believed that major solar eruptions including flares and coronal mass ejections are powered by the free energy stored in the stressed magnetic fields in the so called active regions (ARs). These stressed fields transport magnetic energy and helicity during the evolution of ARs primarily by the mechanisms of flux emergence from sub-photosphere and the foot point shearing motions at the photosphere. Of the many important features, sunspot rotations are a form of uncommonly observed motions, lasting for even days, during the evolution of the ARs \citep{evershed1910,bhatnagar1967, mcintosh1981, brown2003, zhangj2007}, which are suggested to be efficient mechanisms to inject helicity and energy (e.g., \citealt{stenflo1969, barnes1972, amari1996, tokman2002, torok2003}).  

With the increase of observational capabilities both in sensitivity and resolution, sunspot rotation had drawn a considerable attention in an attempt to explain its characteristics in association with the transient activity. A majority of the studies based on observations examined the relationship between the sunspot rotation and coronal consequences \citep{brown2003, tian2006, tian2008}, flare productivity \citep{yanxl2008, zhangy2008,suryanarayana2010}, the association of flares with abnormal rotation rates \citep{hiremath2003, jiangy2012}, non-potential parameters \citep{zhangj2007, kazachenko2009, vemareddy2012b}, helicity injection \citep{vemareddy2012a} etc. 

\begin{figure*}[!htp]
\centering
\includegraphics[width=.97\textwidth,clip=]{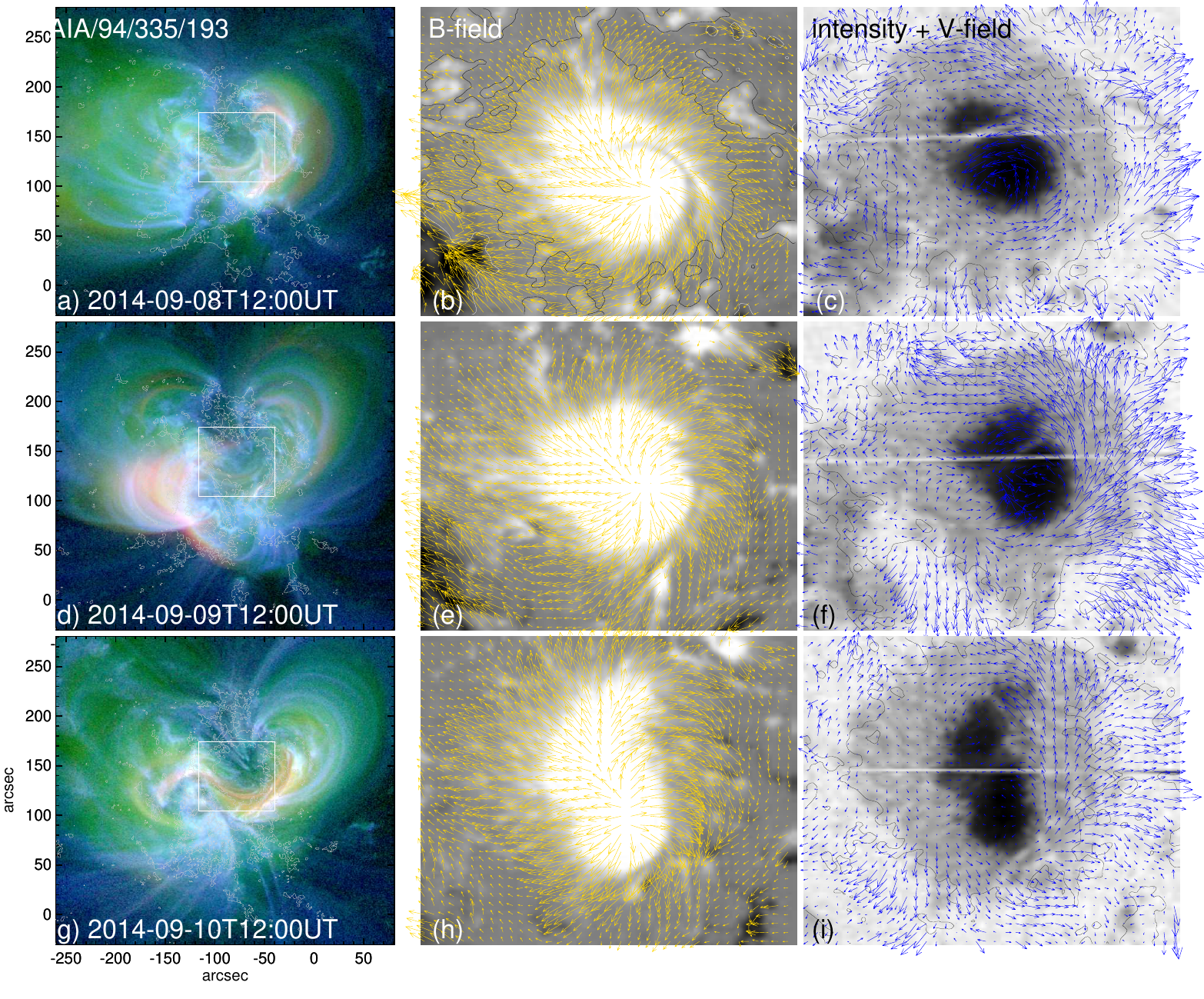}
\caption{Association of sigmoid structure with the sunspot rotation in AR 12158. First column: Snapshots of the corona sigmoid in composite temperatures prepared from AIA 94, 335, 193\AA~channels. Rectangular region indicates the region of rotating sunspot having roots of sigmoid,  Contours of $B_z$ ($\pm150$G) are overlaid to identify the photospheric connections of the sigmoid. Second column: Vector magnetograms of the rectangular region, showing magnetic field in the rotating sunspot. Background is vertical field component and arrows show direction and magnitude of horizontal field. Third column: Horizontal velocity field overplotted on the HMI continuum intensity map of the rotating sunspot. Note the anti-clock orientated velocity vectors on the west portion due to the sunspot rotation. }
\label{Fig1}
\end{figure*}

Numerical MHD investigations also helped greatly in understanding the relationship between sunspot rotation and the eruptive activity by studying the formation and evolution of flux ropes by twisting line-tied potential fields \citep{mikic1990, amari1996, galsgaard1997, garrard2002}. The underlying idea of these simulations is to show that the photospheric vortex motions can twist the core magnetic field in an active region upto a point where equilibrium can no longer be maintained and thus the twisted core field i.e., flux rope, erupts \citep{tokman2002,torok2003, aulanier2010, amari2010}. At the instance of reaching exceeding critical twist, the flux rope is subjected to helical kink instability \citep{torok2005}. Depending on the decay rate of restoring force by overlying field, the progressive injection of the twist in the underlying fluxrope is shown to erupt as a confined flare or a CME. As a secondary possibility, twisting motions could also weaken the stabilizing overlying field of flux rope. Recent numerical model by \citet{torok2013} demonstrates the rotating sunspot as a trigger by inflating the field passing over a pre-existing fluxrope resulting to weaken the downward tension force of the overlying field. In retrospect however, the twisting motions can twist both, the overlying field and the fluxrope, because there is no pure current free field to stabilise the entire flux rope system. Recent observational analysis (e.g., \citealt{vemareddy2014b}) indicates that the kink-instability could be the onset of eruption bringing the fluxrope to a height range of inflating field, from where the eruption is further driven by torus-instability.   

Although the above proof-of-concept simulations strikingly explains and reproduce the many observed features of eruptions, not many observational studies exists to reconcile the developing/formation scenario of flux rope in the host active region of rotating sunspot. In the present paper, we study the developing conditions of sigmoid structure under the influence of non-potential characteristics of rotating sunspot in an active region. Using uninterrupted, high cadence magnetic field observations of AR 11158 at the photosphere, \citet{vemareddy2012b} reported an unambiguous correspondence of sunspot rotation with many non-potential parameters including energy and helicity deposition rates. In that AR, occurrence of the major flares and CMEs are shown to co-temporal with the peak rotation rates of sunspots \citep{jiangy2012, vemareddy2015a}. Importantly, the observed characteristics of those non-potential parameters could have origins of sub-photospheric twist because the AR 11158 was emerging. So for the cause-effect relation, it would be of great interest to investigate a case of sunspot rotation in post phase of AR emergence, which is the subject of this article. Motivated by these studies, we model the AR magnetic structure by non-linear force-free approximations and examined the coronal field topology and current distribution in favour of fluxrope. Observations are outlined in section~\ref{sec2}, results, including measurement of sunspot rotation, non-potential characteristics, and force-free extrapolation are described in section~\ref{sec3}. Summary of the results with a discussion is presented in section~\ref{sec4}.

\section{Observations}
\label{sec2}
The major source of observational data for our study is from Solar Dynamic Observatory. Heliosesmic Magnetic Imager (HMI; \citealt{schou2012}) captures full disc line-of-sight magnetic field measurements at a cadence of 45 s and vector magnetic fields at a cadence of 135s. For the sunspot rotation study, we use continuum intensity observations at 45s cadence. In order to quantify non-potentiality due to the effect of sunspot rotation on the magnetic field, we obtained vector magnetic field measurements at a cadence of 12 minutes provided after a pipelined procedures of inversion and disambiguation \citep{bobra2014, hoeksema2014}. The corresponding coronal activity is studied by multi-thermal EUV images taken by Atmospheric Imaging Assembly (AIA;\citealt{lemen2012}) at a cadence of 12s. 

\begin{figure*}[!ht]
\centering
\includegraphics[width=.97\textwidth,clip=]{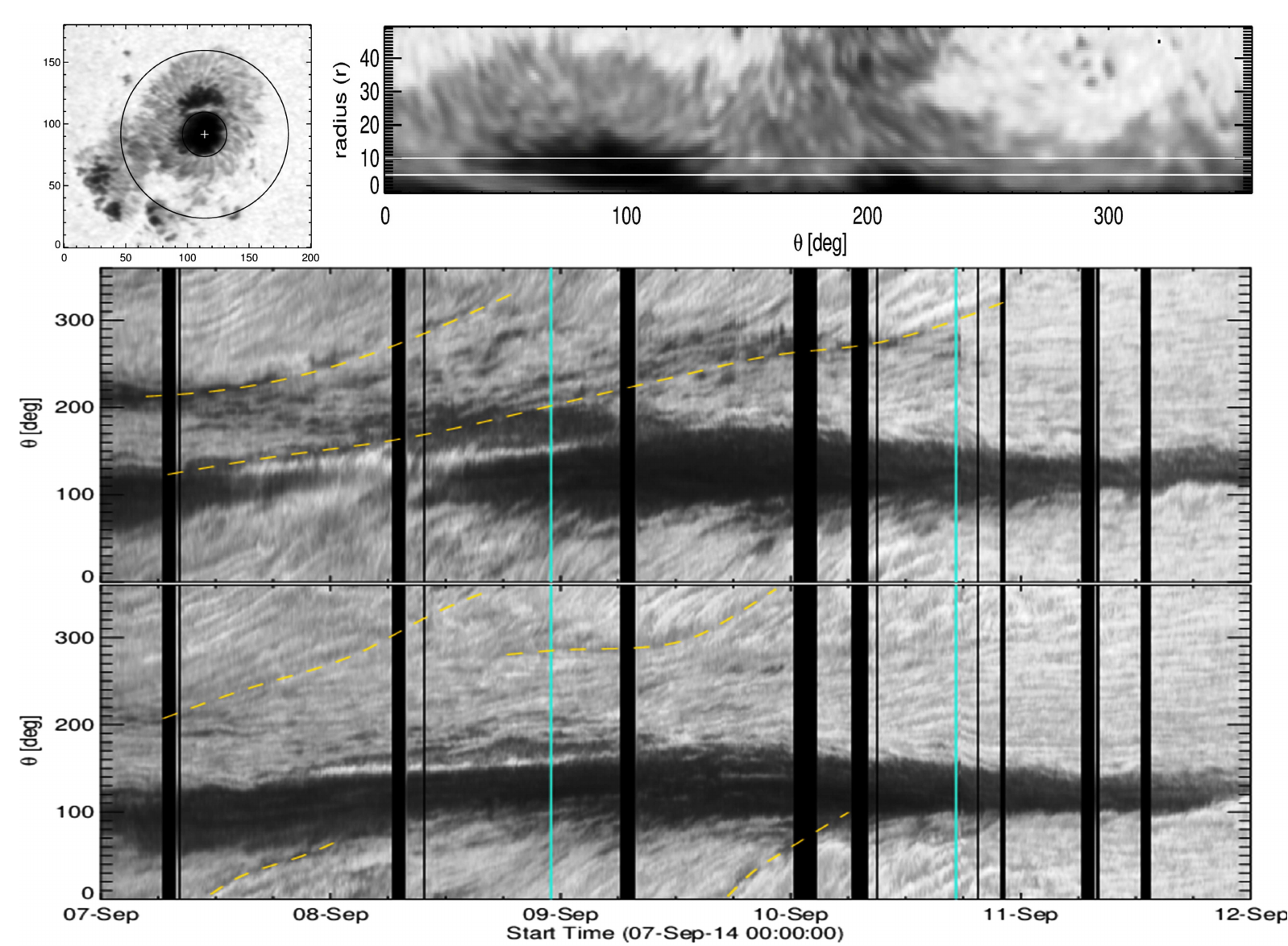}
\caption{Measurement of the sunspot rotation a) Snapshot of the sunspot showing the located centroid ("`+"') and penumbral region between two concentric circles, b) r-$\theta$ map prepared after unstretching the annular region of penumbra of the sunspot. Two horizontal lines indicate the radial positions of the slits to prepare the stack plot of slices,  b-c) Space-time (stack plots) maps at two radial positions from the umbral center. The angular movement of different features refer to the sunspot anti-clock rotation. Dashed curves represent angular motion of some prominent features. Also note that the sunspot features encounter slow/stop down during first half of September 9. Vertical black strips correspond to data gaps and turquoise lines indicate the times of two eruptions.}
\label{Fig2}
\end{figure*}

\begin{figure*}[!ht]
\centering
\includegraphics[width=.7\textwidth,clip=]{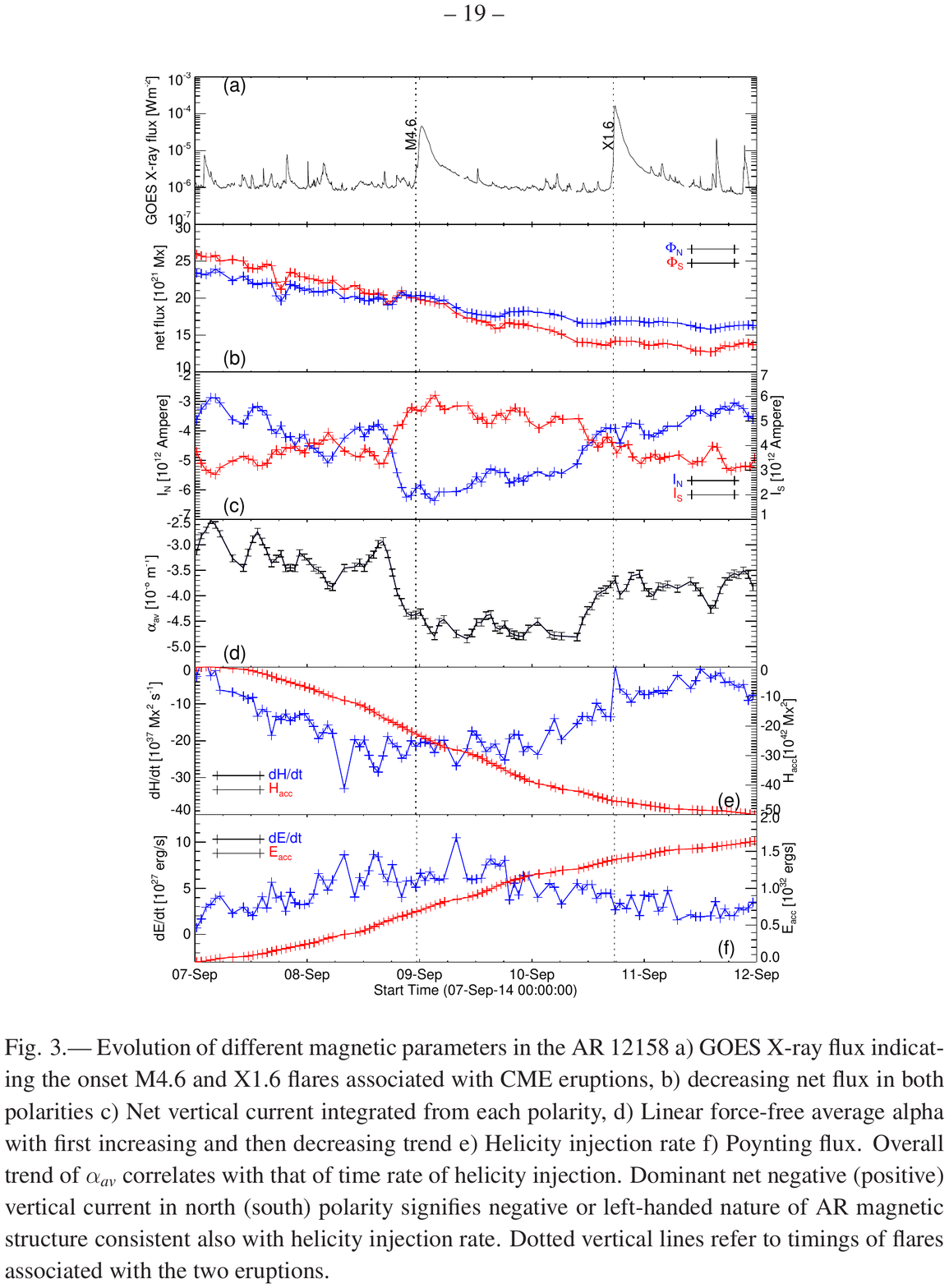}
\caption{Evolution of different magnetic parameters in the AR 12158 a) GOES X-ray flux indicating the onset M4.6 and X1.6 flares associated with CME eruptions, b) decreasing net flux in both polarities, c) Net vertical current integrated from each polarity, d) Linear force-free average alpha with first increasing and then decreasing trend, e) Helicity injection rate, f) Poynting flux. Overall trend of $\alpha_{av}$ correlates with that of time rate of helicity injection. Dominant net negative (positive) vertical current in north (south) polarity signifies negative or left-handed nature of AR magnetic structure consistent also with helicity injection rate. Dotted vertical lines refer to timings of flares associated with the two eruptions. }
\label{Fig3}
\end{figure*}

\section{Results}
\label{sec3}
The active region of interest was NOAA 12158 that appeared on the solar disk during 5-14, September, 2014. It was located on the northern hemisphere at $15^o$ latitude. It is a pre-emerged active region with approximately bipolar magnetic field configuration. During its disk passage, complexity of magnetic configuration ranges from simple $\alpha$ to $\alpha\beta\gamma$. Apart from small scale activity, two major CME eruptions harboured, which are associated with M and X class flares from this AR. The coronal observations captured in multi-wavelengths show a large scale sigmoidal structure. One of its legs have roots from major sunspot of positive polarity. During the time interval of September 7-11, 2014, the composite images prepared from multi-layered observations present multi-thermal plasma loops. During few sigmoid eruptions, these images clearly identifies the presence of hot sigmoid channel surrounded by cool plasma loops (Figure~\ref{Fig1} first column panels). This sigmoid is regarded as a magnetic flux rope \citep{zhangj2012, chengx2013, vemareddy2014b} to connect the theories of flux rope based models to explain the CME eruptions. 

Vector magnetic fields taken by HMI show a main sunspot of positive polarity surrounded by plague type distributed negative polarity. The overall chirality of the transverse vectors aligns in left handed sense. This sense of chirality explains the coronal geometry of magnetic loops which manifests a reverse S-sigmoid (middle column panels of Figure~\ref{Fig1}). Interestingly, the motion images of these vector magnetograms reveal rotation of this main sunspot in anti-clock while the AR evolves persisting to this global sigmoidal structure. To identify the photospheric magnetic connections of the coronal plasma structures, we overlaid contours of magnetic concentrations. They unambiguously show that the sigmoid has roots in that sunspot, indicating that the sunspot rotation has a direct role in progressively building this sigmoidal structure. We used these magnetic field observations to follow the flux motions due to sunspot rotation. By employing the differential affine velocity estimator for vector magnetograms (DAVE4VM; \citealt{schuck2008}), we derived velocity field of the flux motions. In Figure~\ref{Fig1} (last column panels), the horizontal velocity field is overlaid on HMI continuum intensity maps. The orientation of these velocity vectors conspicuously indicate the swirling motion of fluxes owing to sunspot rotation in anti-clockwise direction. Especially, the fluxes from West part of penumbral region exhibits more of this apparent rotation.

\subsection{Measurement of sunspot rotation}
Qualitative measurements of the rotating sunspots are done by preparing the stack plots of a radial section in the penumbra \citep{brown2003, zhangy2008,vemareddy2012b}). The idea essentially is to track the motion of any penumbral feature in time while sunspot rotates about its umbral center (Figure~\ref{Fig2}(a)). For this, the penumbral region is unstretched (anti-clock direction from west) by remapping onto radius-theta plane. We used continuum intensity images from HMI at a cadence of 12 minutes. A snapshot of such an unstretched penumbra in Figure~\ref{Fig2}(a) of our sunspot of interest is shown in Figure~\ref{Fig2}(b).  The stack plots (space-time) are then made by assembling slits taken at a radial position sequentially in time. 
In panels (c) and (d) of Figure~\ref{Fig2}, the stack plots prepared from two different radial positions (5 and 10 pixels from umbra and penumbra boundary) are shown. Since this sunspot rotates in anti-clock direction, we can see the feature motion in increasing angle. The inclination, in time, of penumbral fibrils also delineates a similar physical motion. From these stack plots, we have followed prominently observed feature (dashed yellow curves) motion to derive information about the rate of rotation. Note the feature like white curve is an artifact due to missing data in a row (viz. Figure~\ref{Fig1}(last column)) of intensity image frames. 

From the curves, different features have varying rotation rates in time. Most of the features disappear (or fall into umbra) in a short period of time. Hence it is difficult to track the rotation uniquely with the same feature. On September 7, feature motion is steep followed by slowing motion on September 8. In the first quarter of September 9, the features stagnated in time. This suggests the slowdown of sunspot rotation as also found in the case of AR 11158 \citep{vemareddy2012b, vemareddy2015a}. The magnetic tension in the field lines connecting the sunspot and the opposite negative polarity is suggested to play predominant role in the slowdown of the sunspot rotation. As sunspot rotates slowly, the magnetic stress in the field lines from the sunspot increasingly builds. After a critical point, the tension in these field lines oppose any further rotation. This critical state can also be regarded as non-potential due to stored energy. At this point, any kind of instability may trigger the release of energy. Not surprisingly, a CME eruption is launched at 23:00UT on September 8, followed by M4.6 flare at 23:12UT. In AR11158 also, at the time  (18:00UT on February 14, 2011) of slowdown of sunspot rotation, a major CME eruption triggered followed by M2.2 flare. This observation is a direct consequence of often observed sunspot rotation in horboring the powerful CMEs. The cause of sunspot rotation is likely related to sub-photospheric dynamics, which cannot be probed by photospheric observations. 

Once the tension in the connected field lines released by an eruption, the sunspot rotation may continue depending on the driving force beneath the photosphere. Here in our case, after first eruption, the sunspot rotates till September 11. The feature motion is ambiguous to infer any further cessation of rotation, however another powerful eruption launched at 17:15UT on September 10, followed by an X1.6 flare at 17:21UT. From these observational cases, we suggest that the slowdown of the sunspot rotation is an indicator of a triggering major powerful eruptions.

From the time profiles of the rotation, we derived rotation rates ($d\theta /dt$) of different features. We found them to rotate at varying rates. A feature that retains long time have $d\theta/dt$ of 0-4 deg/h. However, there are short lived features (6-8h, in 0-100deg section) that rotate even fast upto 8deg/h. In all, sunspot rotation is neither uniform in around it nor constant over the time.

\begin{figure*}[!ht]
\centering
\includegraphics[width=.97\textwidth,clip=]{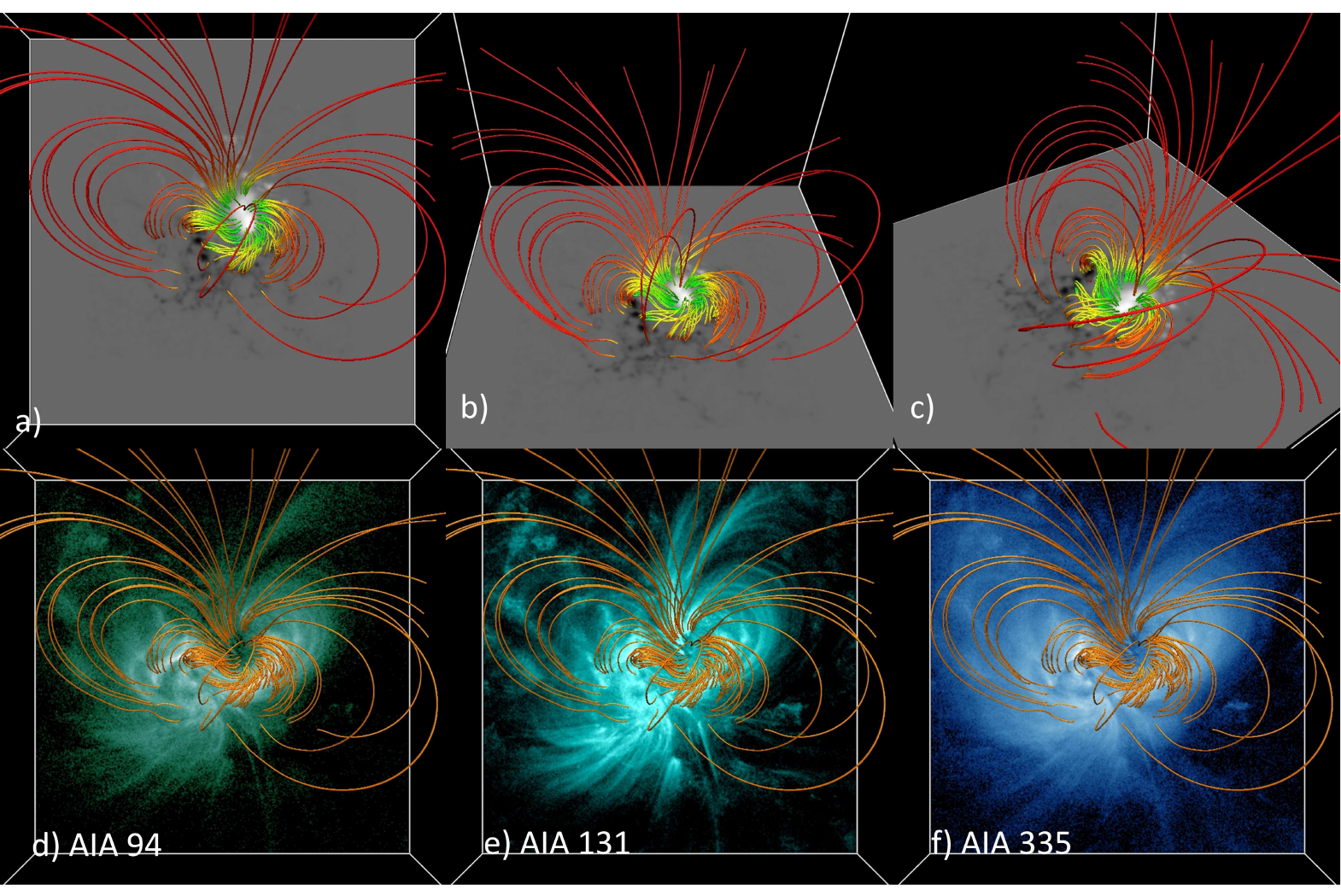}
\caption{Magnetic structure in AR 12158 at 12:00UT on September 10, 2014. (a-c) Traced field lines imaged over $B_z$. Field lines from the periphery of the sunspot manifest the body of the sigmoid and those from interior overly the sigmoid.  d-f) Field structure on the AIA images reflecting the resemblance of a flux rope structure.}
\label{Fig4}
\end{figure*}

\subsection{Evolution of magnetic non-potential parameters}
Under the presence of unusual rotating sunspot in the active region, it is imperative to study the evolution of magnetic non-potentiality. It is quantified by many parameters, but few of them that can be estimated with the photospheric vector magnetograms ({\bf B}) are net vertical current,$\alpha_{av}$,  helicity injection, and poynting flux etc. Vertical current distribution is calculated by 
\begin{equation}
{{j}_{z}}=\frac{\left( \nabla \times \mathbf{B} \right)}{{{\mu }_{0}}}
\end{equation} 
Where ${{\mu }_{0}}=4\pi \times {{10}^{-7}}Henry\,{{m}^{-1}}$. The extent of averaged twistedness of magnetic structure in the AR is estimated by 

\begin{equation}
{{\alpha }_{av}}=\frac{\sum{{{J}_{z}}(x,y) sign[{{B}_{z}}(x,y)]}}{\sum{|{{B}_{z}}|}}
\end{equation}
\citep{hagino2004}.  The sign of this parameter generally gives the handedness or chirality of the magnetic field. Helicity injection rate relates the flux motions with the observed twisted magnetic field \citep{berger1984} by
\begin{equation}
{{\left. \frac{dH}{dt} \right|}_{S}}=2\int\limits_{S}{\left( {{\mathbf{A}}_{P}}\bullet {{\mathbf{B}}_{t}} \right){{\text{V}}_{\bot n}}dS}-2\int\limits_{S}{\left( {{\mathbf{A}}_{P}}\bullet {{\mathbf{V}}_{\bot t}} \right){{\text{B}}_{n}}dS}
\end{equation}
 where ${\bf A}_p$ is the vector potential of the potential field ${\bf B}_p$, ${\bf B}_t$ and $B_n$ denote the tangential and normal magnetic fields, and ${\bf V}_{\bot t}$ and $V_{\bot n}$ are the tangential and normal components of velocity $V_{\bot}$, the velocity perpendicular to the magnetic field lines. The velocity field ({\bf V}) is derived from time sequence vector magnetic field observations obtained HMI by employing DAVE4VM technique. Similarly, the magnetic energy (Poynting) flux across the surface \citep{kusano2002}, can be estimated as

\begin{equation}
{{\left. \frac{dE}{dt} \right|}_{S}}=\frac{1}{4\pi }\int\limits_{S}{B_{t}^{2}{{V}_{\bot n}}dS-\frac{1}{4\pi }}\int\limits_{S}{\left( {{\mathbf{B}}_{t}}\bullet {{\mathbf{V}}_{\bot t}} \right){{B}_{n}}dS}
\end{equation}
Procedures involving the estimation of these parameters are widely described in many recent studies (e.g., \citealt{vemareddy2012b, vemareddy2012a, liuy2012, vemareddy2015b}) in different contexts. On following similar procedures, we calculated these parameters in this AR and plotted their time evolution in Figure 3. The net flux from north and south polarity shows monotonous decrease from the start of the observation interval. Imbalance of the flux content in the AR is in the range of 7-11\%. Net vertical current in the north ($I_N$) polarity is negative and varies from $-3\times10^{12}$A to $-6.5\times10^{12}$A. On the other hand, it is positive in the south polarity ($I_S$) varying from $3\times10^{12}$A to $6\times10^{12}$A. As the sunspot keep rotates, the shear in horizontal vectors keep increases, which in turn contributes to the net vertical currents in the form of horizontal field gradients. As both of these currents have reached their maximum values by the time of cessation of the sunspot rotation. As per these current profiles, $\alpha_{av}$  also shows first increasing trend upto $-4.7\times10^{-8}m^{-1}$ till the time of the first eruption and then decreases followed by major eruption at 17:30UT on September 10.  

\begin{figure*}[!ht]
\centering
\includegraphics[width=.97\textwidth,clip=]{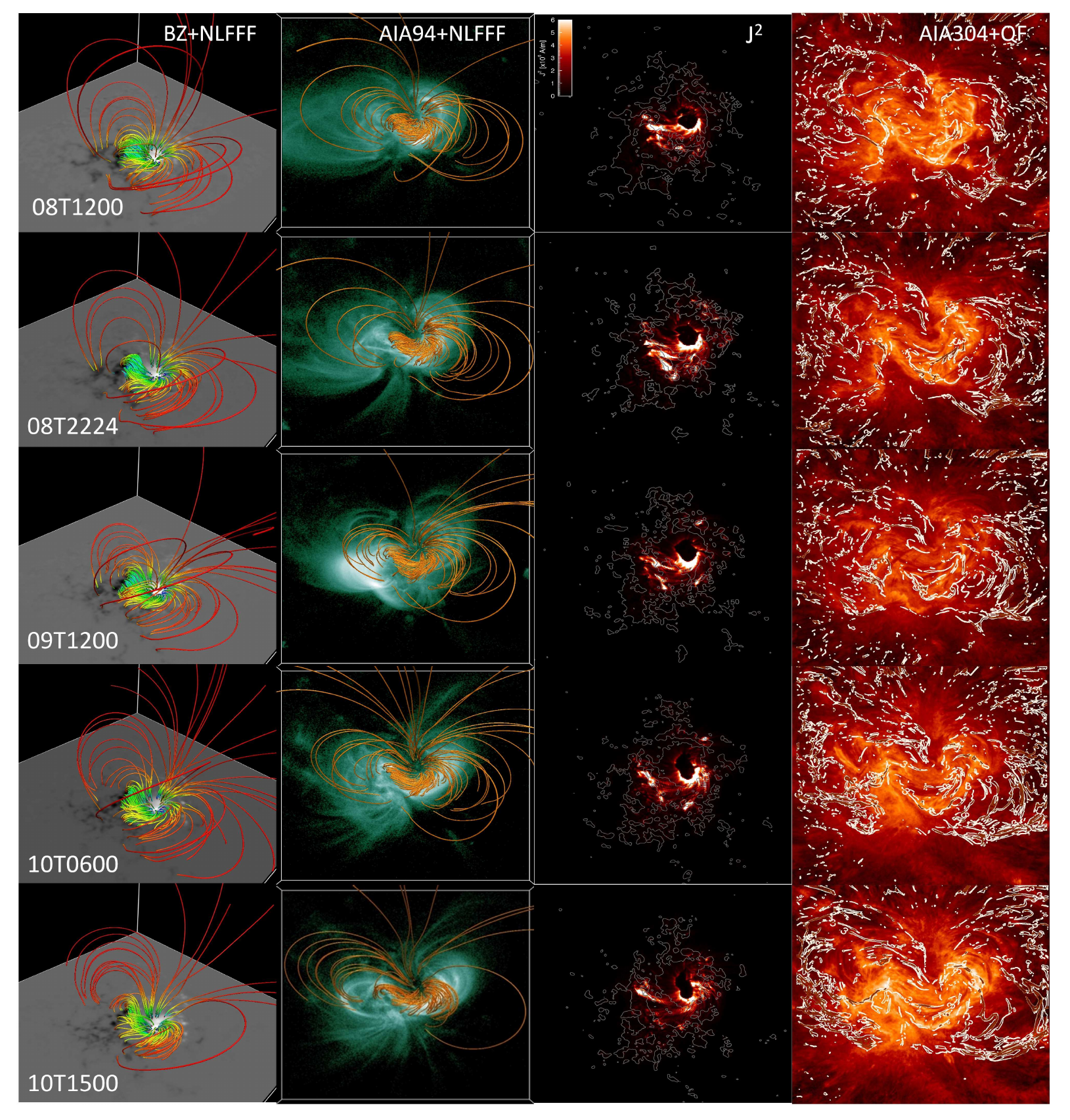}
\caption{Magnetic structure around the rotating sunspot, from top to bottom, at different epochs. First Column: Top view of the extrapolated field structure, resembling the sigmoid. Second column: Tilted view of the same to visualize the stressed nature of the field lines around the sunspot Third column: Vertical integrated maps of current density $|{\bf J}^2|$ to locate the build up of strong current concentrations. Fourth column: AIA 304\AA~images overlaid by contours of QF. EUV emission along the sigmoid outlines the higher QF values and indicates co-spatial location of highly stressed magnetic field, current concentration.}
\label{Fig5}
\end{figure*}

Moreover, the time profile of $dH/dt$ also exhibits a similar trend as the net current and $\alpha_{av}$. The order of the estimated values of $dH/dt$ and $\Delta H$ are consistent with the earlier studies (e.g., \citep{vemareddy2012a, vemareddy2012b}). It is worthwhile to point that the dominant net negative (positive) vertical current in north (south) polarity signifies a negative or left handed nature of AR magnetic structure consistent also with the helicity injection rate. All these profiles suggest that the non-potentiality is a direct consequence of organized flux motions generated by the sunspot rotation. Energy flux injection is positive and is on the orders of 27 ergs per second. The accumulated energy over the time interval before the occurrence of the eruption and the associated flare is of the orders of 32 ergs, that is suffice to generate a flare of magnitude upto GOES class X. A similar observational results follows from the study of sunspot rotation in AR 11158 \citep{vemareddy2012a, vemareddy2012b}, where peak phase (in magnitude) of different non-potential parameters coincides with the occurrence of the major eruptions and associated flares. This demonstrates that the successive accumulation of non-potentiality is mainly due to the surface motion (shear/twist) on the photosphere but not the flux emergence.

\subsection{Non-Linear Force-Free modelling}

In order to realise the effect of sunspot rotation on the geometry of the AR magnetic structure, we performed non-linear force free field (NLFFF) extrapolation \citep{wiegelmann2004, wiegelmann2010} of the observed photospheric magnetic field. The field-of-view of the boundary field covers full AR such that flux is nearly balanced during the entire time interval. To satisfy the force-free conditions, the magnetic components are subject to the pre-processing procedure \citep{wiegelmann2006}. To facilitate tracing field lines in a large extent of volume, the observed boundary is inserted in an extended field-of-view and computations are performed on a uniformly spaced computational grid of $400\times400\times256$ representing a physical dimensions of $292\times292\times280$ Mm$^3$. Before this, we rebinned the observations to 1 arcsec/pixel. The NLFFF code is initiated with linear force-free field constructed from normal field component and a small value of force-free parameter.

\begin{figure*}[!ht]
\centering
\includegraphics[width=.97\textwidth,clip=]{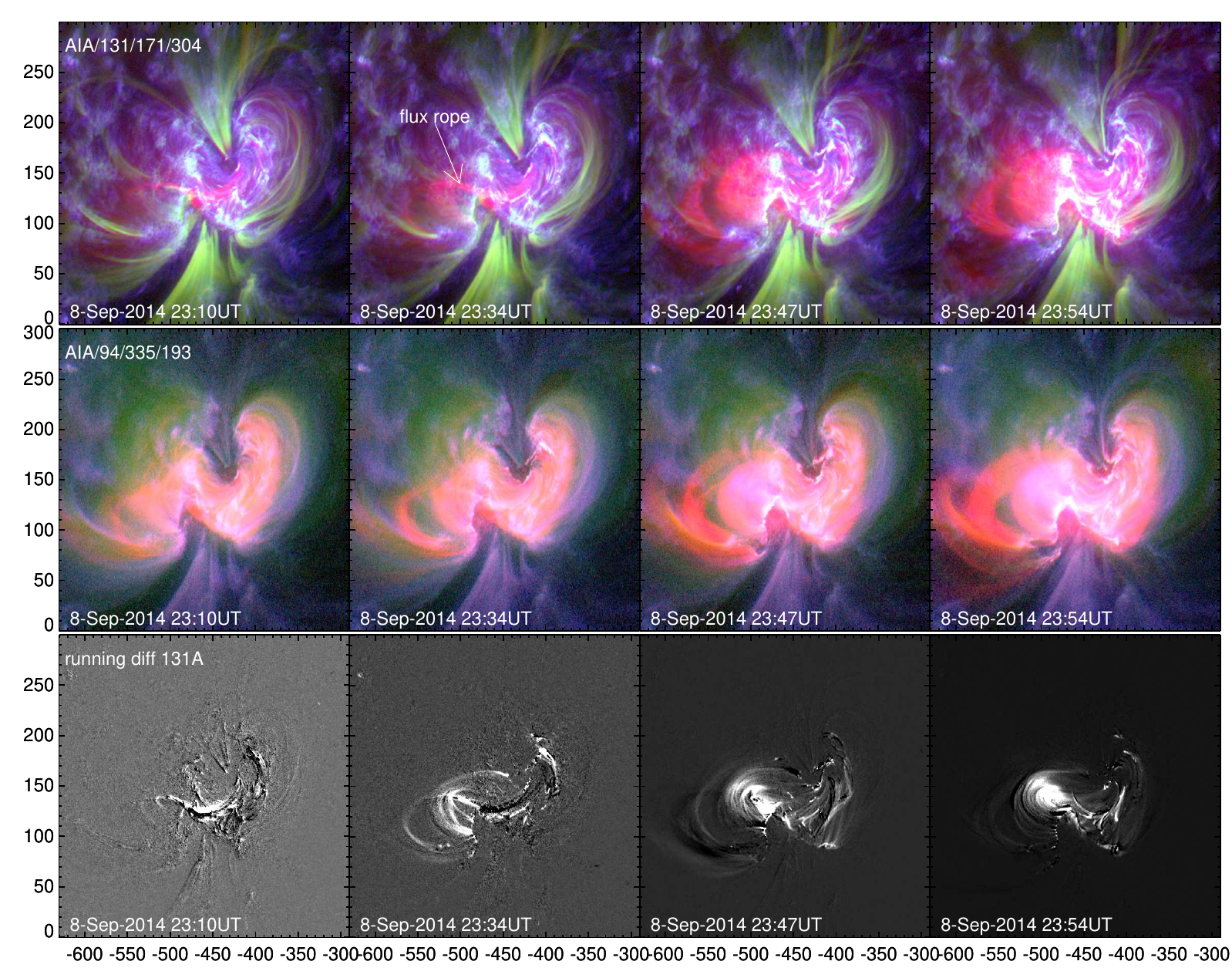}
\caption{Onset scenario of sigmoid eruption on September 8, 2014 at around 23:55UT. {\bf Top row:} Composite images of corona observed in different temperatures of AIA 131, 171 and 304\AA~channels. Note the flux rope as a hot continuous trace during its onset of slow rise motion {\bf Middle row:}  Composite images of the corona in another combination of AIA temperatures channels (AIA/94/335/193). Note the enhancing hot flux bundle (flux rope) in the middle of the sigmoid structure.  {\bf Bottom row:} Running difference images of the AIA 131\AA~showing the rising sigmoidal structure before the onset of the coronal mass ejection. The timing of this eruption precedes the drastic increase of net vertical current and $\alpha_{av}$. All panels are in heliographic arcsec units.}
\label{Fig6}
\end{figure*}

With the extrapolated field in the coronal volume above the AR, we traced field lines roughly according to the total current (|{\bf J}|) and total field strength (|{\bf B}|). This modelled structure is plotted in Figure~\ref{Fig4} with $B_z$ (panels a-c) and coronal EUV observations (panels d-f) as the bottom boundary. Field lines from the lower periphery of the sunspot correspond to the middle section (body) of the sigmoid whereas those from central part serve as overlying flux system. Right J-section of the sigmoid is highly curved due to more twisted field lines from the top periphery of the sunspot, where the modelled structure differs significantly. However, the field lines closely resemble the global magnetic structure of plasma loops of the sigmoid and surrounding loops as the earlier modeling studies \citep{savcheva2009, xudong2012}. As sunspot rotates in anti-clockwise direction, the field lines tend to retain their connectivity, and appear as swirled in clockwise direction. As a fact of high current density, the flux system between the sunspot and its counterpart appears as hot channel when observed in coronal AIA 131, 94\AA~wavelengths. 

Now the entire magnetic system is evolving quasi statically, because the driving boundary motion (~1km/s) is far less than the Alfven time scale of the coronal magnetic field. Therefore, our static modelling cannot capture the features of rapid evolution during sigmoid eruption. However, the gradual build up, like flux rope current channel, topology which are the basic building blocks of eruption models \citep{torok2005, aulanier2010}, can approximately be captured. A close view of the magnetic structure around the sunspot reveals the effect of sunspot rotation as described in the earlier sections (panel (c)). Due to this twisting motion especially at the sunspot periphery, the field lines rooted therein had a fan-shaped structure (deviating from radial ones), the field lines rooted near to the center overly the earlier. The two J-sections of the sigmoid are compact and curved which the NLFFF code fails to reproduce exactly. 

For a view of quasi-static evolution, in Figure~\ref{Fig5}(columns 1 and 2), we plotted the magnetic structure in the sigmoid at different epochs of sunspot rotation. To capture the sheared field lines around the sunspot, field lines are rendered according to higher current density criteria. Highly sheared field lines originate from the lower periphery of the sunspot and they lie mostly below 10Mm in height. The two J-sections of the sigmoid are compact and curved which the NLFFF code fails to reproduce exactly. In such a flux system being slowly driven by constant rotational motion, the formation of the fluxrope under the dynamic scenario is inevitable. Moreover, it appears that continuous sunspot rotation helps sustain the fluxrope structure throughout the evolution.

The current density characterises the non-potentiality of the field. The patterns of strong current concentrations serve as a proxy to non-potential structure in the corona. Moreover, current structures are regions where reconnection can occur to convert magnetic energy to thermal and kinetic energy.  Dense distribution of current persists mostly around lower portion of the sunspot upto a height of 10Mm. This immense coronal current distribution is due to increasingly developing sheared arcade interfacing the rotating sunspot and the surrounding negative polarity at the lower half portion. We compute the vertical integration of $J^2$ (i.e., $\int\limits_{z}{{{J}^{2}}dz}$) (column 3 in Figure~\ref{Fig5}). As $J^2$ term is proportional to the Joule heating term, it thus roughly represents the hot emission. This is indeed true in our case. The strong current concentration around the sunspot spatially coincides with the high intensity of EUV emission in 304~\AA~images, especially south circular portion due to the highly stressed magnetic field. 

To measure the magnetic field line linkage, we also compute the quashing factor QF \citep{titov2002}. Higher value of QF locates the quasi-separatrix layers which are the sources of high current concentrations. The contours of QF on AIA 304\AA~observations are shown in Figure~\ref{Fig5} (fourth column panels). The traces of high QF-values roughly outline the sigmoid in all the time shots. The difficulty of reproducing curved, compact J-sections is well acknowledged due to not-well enough observational sensitivity of HMI \citep{nindos2012, xudong2012, vemareddy2014a}. Owing to this difficulty, reproducing a flux rope structure with the extrapolation technique deemed to be a challenge and different treatments to the boundary observations (e.g., \citealt{jiangc2012, jiangc2014, zhaoj2016}) are being employed in different extrapolation codes. The optimization code relies on observed horizontal field components and globally relaxes toward force-free equilibrium. Without any treatment, the model remarkably shows many similarities of the flux rope structure around the rotating sunspot.  In the following, we explored the evolution of the magnetic structure around sunspot over the time of two observed major eruptions. 

\begin{figure*}[!ht]
\centering
\includegraphics[width=.99\textwidth,clip=]{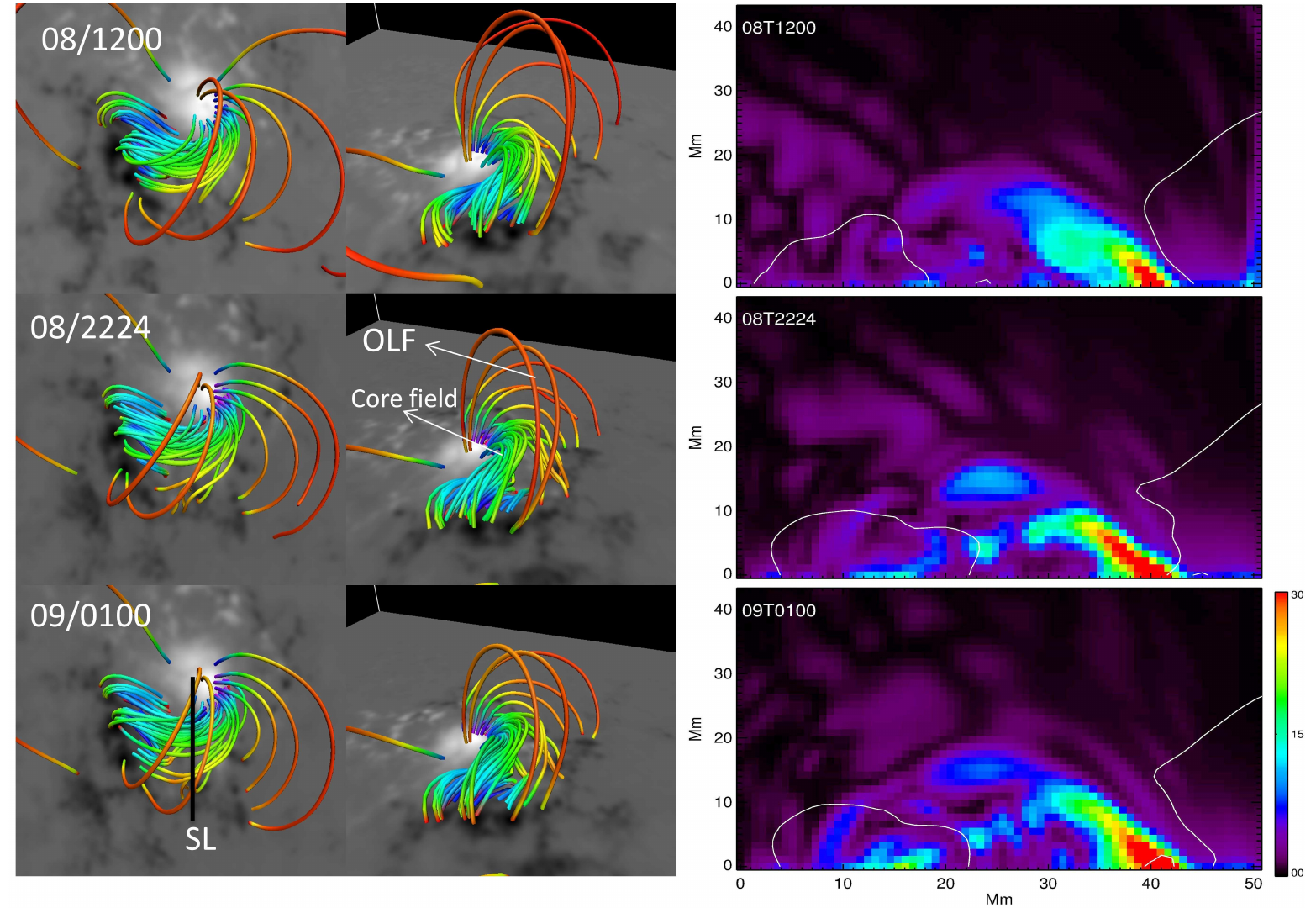}
\caption{Scenario of flux rope development from the sheared arcade around the sigmoid eruption event of September 8, at 23:55UT. {\bf First column:} Top view of the field structure rendered from the rotating portion of the sunspot.  Straight line SL refers to vertical slice to examine the current distribution across the sigmoid structure. {\bf Second column:} Side view of the field structure in the corresponding panels of first column.  It indicates the flux rope structure with twisted core and overlying potential field envelope. {\bf Third column:} Current distribution $|{\bf J}|$ ($mAm^{-2}$) in the vertical slice SL shown in the corresponding panels of first column. White curves are contours of $B_z$ obtained in the same SL. Note the strong current concentrations located low in the height overlaid by weak current concentrations of less sheared arcade.}
\label{Fig7}
\end{figure*}

\begin{figure}[!ht]
\centering
\includegraphics[width=.49\textwidth,clip=]{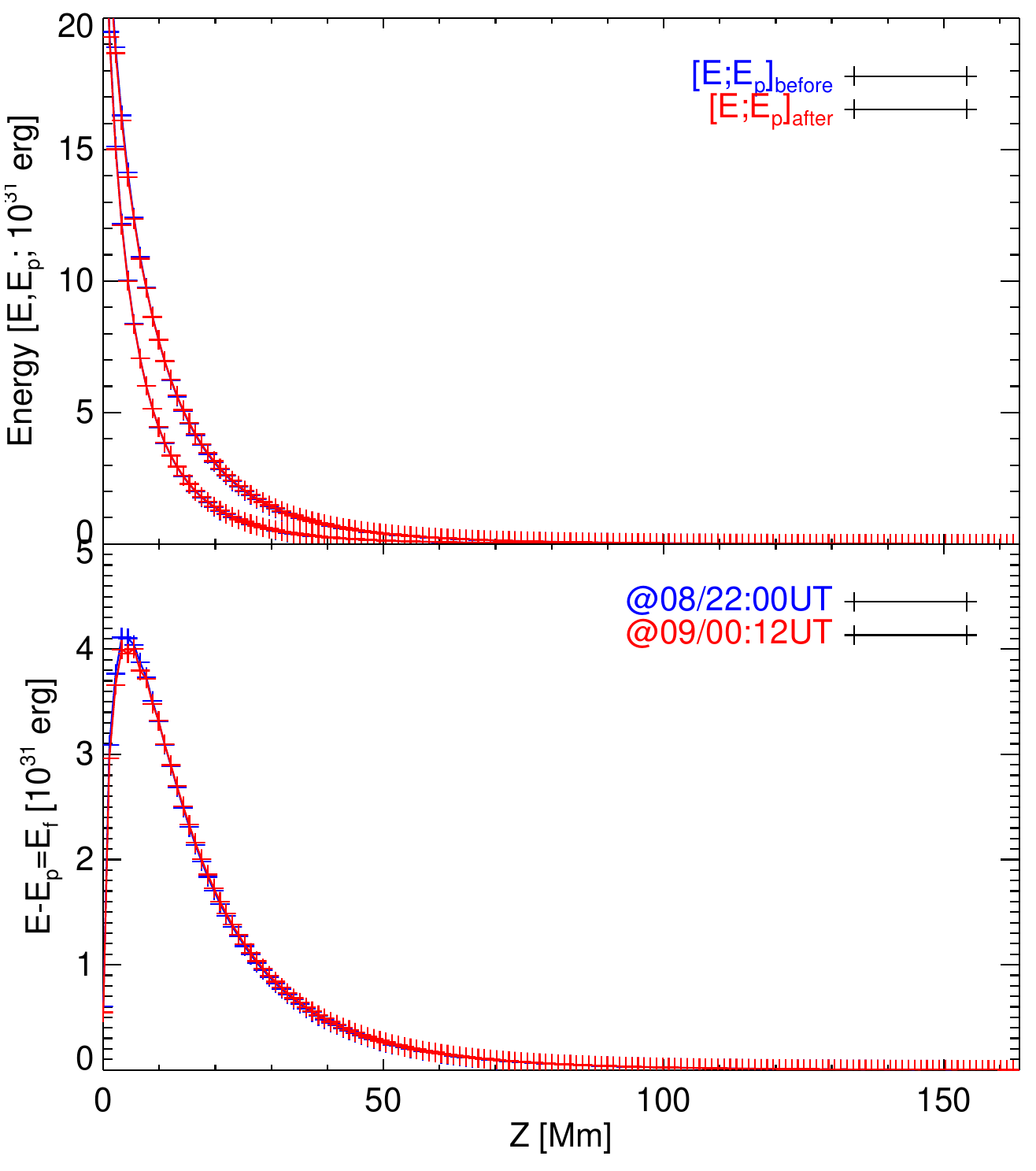}
\caption{Evolution of magnetic energy before and after the eruption event of September 8, 2014 in the AR12158. {\bf top} Non-potential energy density and potential energy density are integrated over each horizontal plane and plotted as a function of height. {\bf bottom} Horizontally integrated free energy, as a function of height, before and after the eruption (bottom). }
\label{Fig8}
\end{figure}

\subsection{Eruption event of September 8, 2014}
A moderate eruption occurred in this AR at 23:50UT when the AR is at disc location of E30N15. It generates a major disturbance in the corona as a CME and M4.6 flare. Figure 6 illustrates the dynamical scenario of this eruption event from high cadence multichannel AIA observations. For a simultaneous view, two combinations of composites are considered in this study. Composites prepared by AIA 94/335/193 channels (middle row panels) clearly present a well-developed structure of sigmoid just before the main eruption. On the other hand, the composites prepared by AIA 131/171/304\AA~(top row panels) shows a rising continuous flux thread (as a main body of flux rope) in the middle of the sigmoid channel. This is also clear from difference images of 131\AA~where very hot flux rope channel essentially captured (bottom row panels). This is consistent with recently settled debate on time-instance of formation and appearance of the flux rope \citep{zhangj2012, chengx2013, vemareddy2014b} in the solar source regions. Here, the sigmoid structure (regarded as flux rope) preformed preferably by the continuous action of sunspot rotation, and we see its existence as a continuous flux bundle (embedded in sheared arcade) only during initiation (around 23:00UT, top row panels) of its slow rise motion in hot channels. 

Since the flux rope is having magnetic connections with the rotating part of the sunspot, kink-instability is likely to involve in the onset of the flux rope eruption by constantly injecting twist into the flux system constituting the main body of the sigmoid. This can be checked by relating $\alpha_{av}$ to twist number in the coronal loop constituting the sigmoid \citep{leamon2003}. Total twist T of the coronal magnetic loop, assuming as a semicircle of length l with its footpoint separation distance d is given by
\begin{equation}
T=lq=\frac{\pi d}{2}\frac{\alpha_{av}}{2}=l\frac{{{\alpha }_{av}}}{2}
\end{equation}
Here, winding rate is assumed as half the  $\alpha_{av}$ because it is not a well-known parameter. Since the local values of  $\alpha=\frac{J_z}{B_z}$ are of order $10^{-6}m^{-1}$, averaging over small area at flux rope leg in the sunspot region gives $>0.7\times10^{-7}m^{-1}$. As the traced sigmoid length is about 190Mm, the above expression implies total twist of more than one turn (6.65 radians=1.05 turns, note a $2\pi$ factor with turns). Note that high resolution and high cadence observations may improve the calculations in which case the $\alpha$ value, on average, may indicate kink-nature of field lines constituting the flux rope. Reconnection with the overlying field lines \citep{antiochos1999} in a later phase triggers the eventual eruption of this FR at 23:50UT, which follows the commencement of M4.6 flare. Unlike usual cases, the progressive reconnection during post-flare phase lasts for 10hours. Even after this long duration flare event, the precursor sigmoid structure retains its geometry, indicating the eruption as a partial one \citep{gilbert2007}. The associated CME was captured in LASCO/C3 field-of-view and found to have a linear speed of 230km/s\footnote{\url{http://cdaw.gsfc.nasa.gov/CME_list/UNIVERSAL/2014_09/}}.

\begin{figure*}[!ht]
\centering
\includegraphics[width=.97\textwidth,clip=]{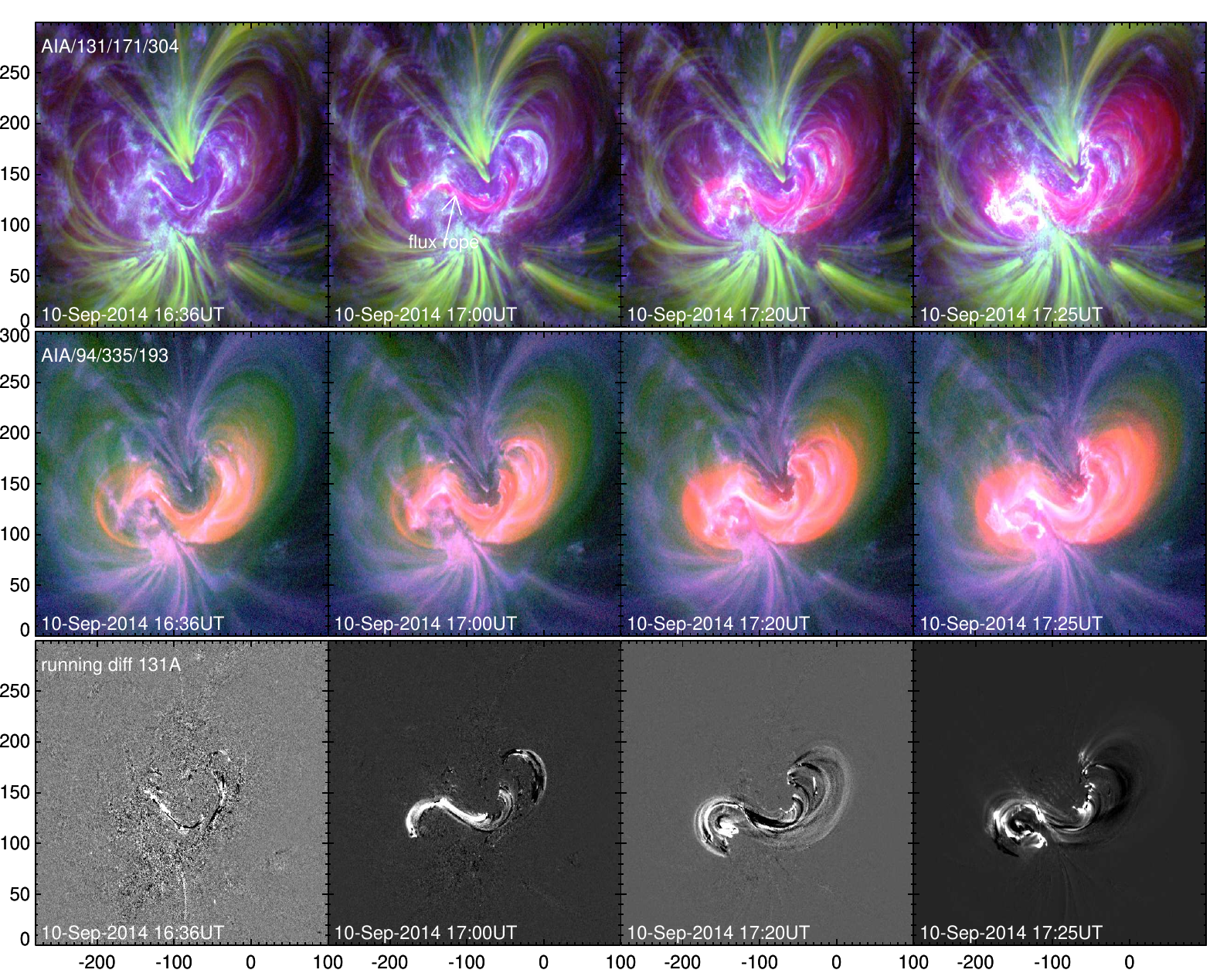}
\caption{Same as Figure~\ref{Fig6} but for the onset of sigmoid eruption of 17:30UT September 10, 2014. This eruption leads to an Earth directed fast coronal mass ejection.}
\label{Fig9}
\end{figure*}

Visualization of the field lines from rotating portion in the sunspot suggests the development of the fluxrope structure. In Figure~\ref{Fig7}, we plotted those field lines for the snapshots around this eruption event.  The lower field lines are progressively sheared by the sunspot rotation and become twisted core of the overlying less sheared field lines. We then compute the current density in a vertical cross-section plane (slice SL) of this fluxrope structure and plotted in the corresponding panels of third column. Distribution of $B_z$ in the same slice planes is also computed and over plotted its contours ($\pm500$G) to identify the main polarities. Owing to stressed field lines all along the polarity inversion line of the sunspot and the negative polarity, the current distribution appears exactly in a arcade shape and strong current concentration is co-spatial with the polarity inversion line. With the development of the twisted core structure, the current concentration above the negative polarity enhanced (22:00UT panel). This is obvious even from the time profile of the net vertical current (Figure~\ref{Fig3}), where surface integral of vertical current density increases rapidly (over a span of 5h) during this eruption event. We emphasize the difficulty of capturing the twisted or helical structures at the sigmoid sites due to lack of sufficient instrumental sensitivity for horizontal polarisation signal \citep{hoeksema2014} and the model of boundary dependent extrapolation as well. This issue deserves a separate study with a different treatment of data driven simulation (e.g.,  \citealt{wust2006, jiangc2014}) and would be our future investigation. 

It is a matter of interest to estimate the total magnetic energy ($E=\int\limits_{V}{\frac{{{B}^{2}}}{8\pi }d\text{V}}$) of the AR magnetic system under these evolving conditions. In addition to global energy content, we can also study the height variation of the magnetic free energy. For this, we computed the surface integral of the magnetic energy 
\begin{equation}
E(z)=\int\limits_{S}{\frac{{{B}^{2}}}{8\pi }}dxdy
\end{equation}

which tells the height dependent energy content (e.g., \citealt{mackay2011}). Using this expression, we compute potential field energy ($E_p$), total non-potential energy (E), and free magnetic energy ($E_{free}=E- E_p$) before and after the enhancement of the total current (.i.e., eruption event) in the AR magnetic structure. They are plotted in Figure~\ref{Fig8} with respect to height from photosphere into the corona. We can see from these height profiles that E and $E_p$ are predominantly located below 20Mm. Although they minutely differ from one another, the $E_f$ curves well distinguished. $E_f$ is mainly situated in the height range of 2Mm to 40Mm with a maximum at about 6Mm.  

\begin{figure*}[!ht]
\centering
\includegraphics[width=.95\textwidth,clip=]{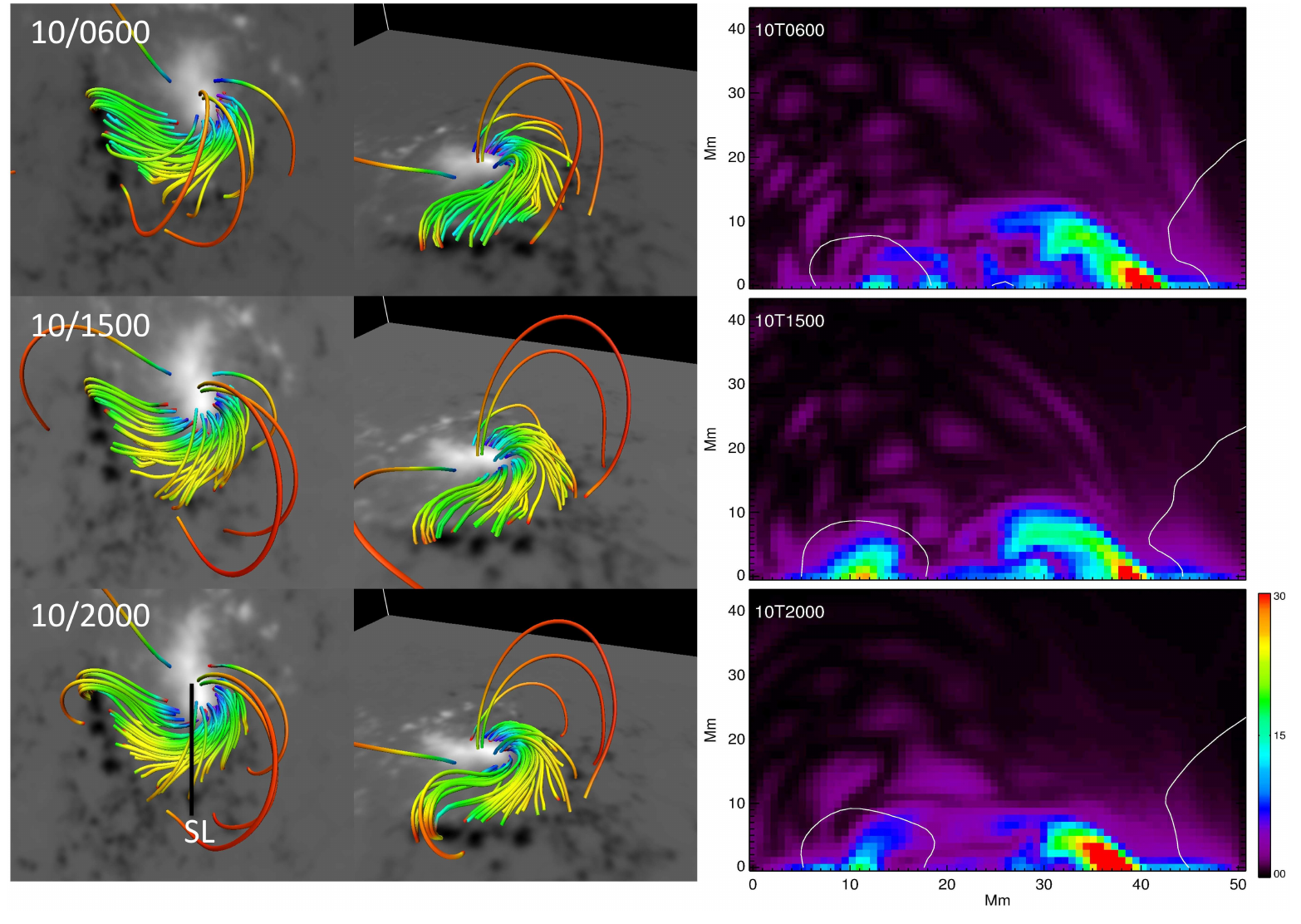}
\caption{Same as Figure~\ref{Fig7}~but during the sigmoid eruption event of 17:30UT September, 2014. In all three slices, total current density smears away probably due to dissipation by reconnection.}
\label{Fig10}
\end{figure*}

\begin{figure}[!ht]
\centering
\includegraphics[width=.49\textwidth,clip=]{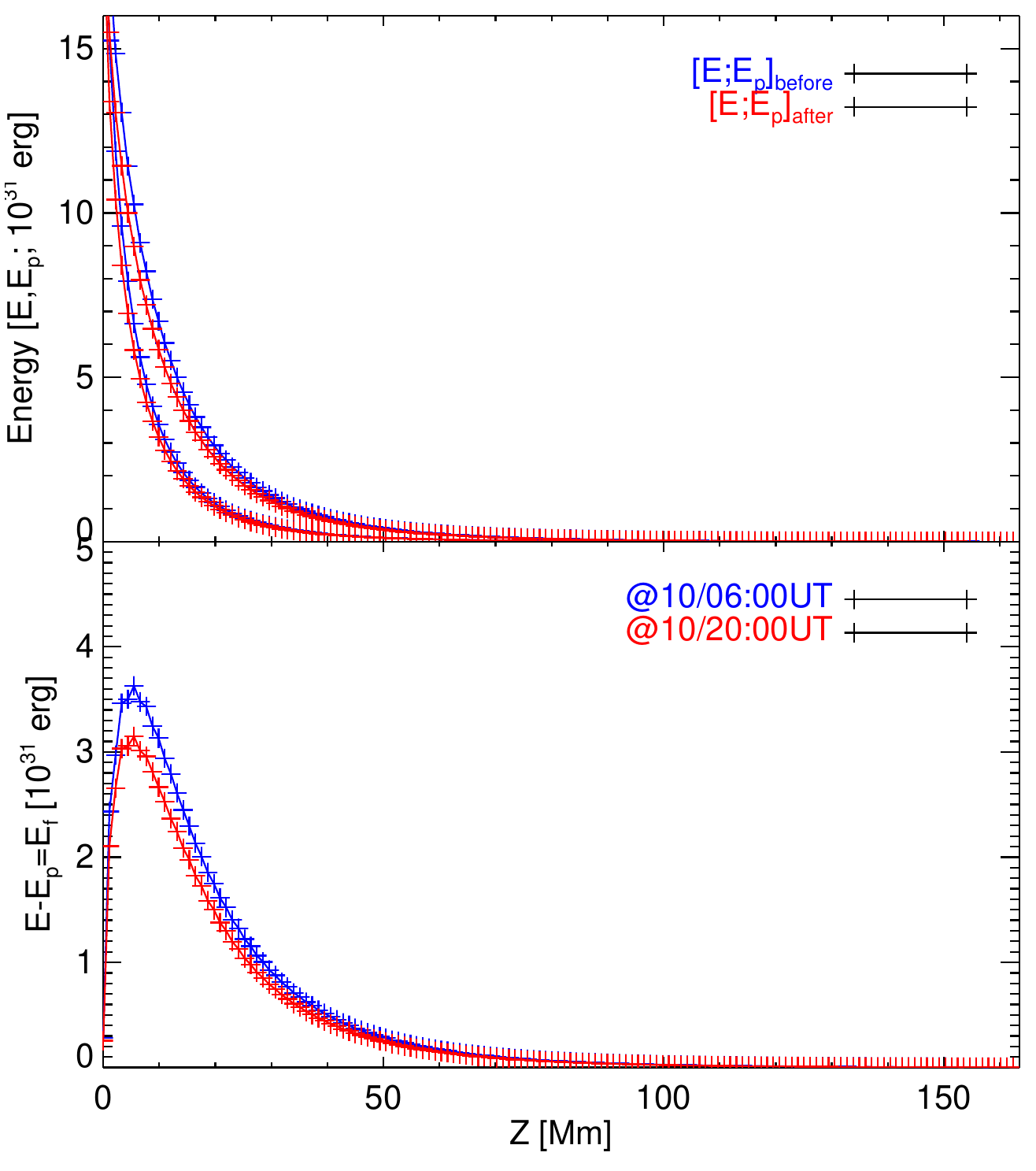}
\caption{Same a Figure~\ref{Fig8} but for the sigmoid eruption event of 17:30UT  September 10, 2014 in the AR12158.}
\label{Fig11}
\end{figure}

From these profiles, we can calculate the global energy loss/gain during this eruption. We found the free energy before (22:00UT on September 8) and after (00:12UT on September 9) as $79.8\times10^{31}$erg, $79.7\times10^{31}$erg respectively. The free magnetic energy that would be available for this event is positive ($0.1\times10^{31}$erg). Although it is marginally sufficient to power the observed flare,  the quantity is still small for an eruption. The reason lies in the fact that the observed field components showed increased net vertical current around this event, however snapshots just before and after this event detects available free energy. Moreover, the required energy for this M-class flare releases from a localised region by field reconfiguration and the averaging over entire volume may not detect it over globally building scenario. Since the eruption is associated with low speed CME, it is likely that the released energy is small.    

\subsection{Eruption event of September 10, 2014}
A second major eruption from AR 12158 occurred on September 10, 2014 at 17:25UT. Since the AR is at disc location of E5N15, the CME eruption is face-on to the Earth and a halo CME at a linear speed of 1071km/s is registered in LASCO white-light CME observations. Unlike the earlier one, this event is a full successful eruption distinguished by the speed of the CME and the nature of speed of reconnection associated with flare which is X1.6. In Figure~\ref{Fig9}, we displayed the snapshots of the coronal imaging observations during the onset of the eruption. The eruption commences from 16:45UT, since then the visibility of continuous trace of flux rope becomes apparent in the composite images of 131/171/304~\AA. Increasingly enhanced emission all along the sigmoid is likely be a consequence of reconnection in a thinning current sheet that would form below the flux rope (\citealt{gibson2006}, second row panels of 94/335/193 composites). Like the earlier event, this event is also suggested to be a consequence of helical kink instability triggered under the continuous slow driven motion by the sunspot rotation. The evidence comes from the analysis of localized distribution of $\alpha$ as described for the previous event.

The field structure seen in the snapshots around this event also suggest a flux rope (Figure~\ref{Fig10}). The core part is not as strongly twisted as the earlier case but overlaid by sheared arcade. Strong current concentrations due to sheared arcade system of sigmoid are distributed upto the height of 15Mm well before this eruption. These current concentrations are located along and above the PIL owing to the stressed field between the sunspot and its surrounding opposite polarity. Moreover, the appearance of fluxrope in EUV channels is only during the onset time, which is too highly dynamic to be followed by the static models based on observed static photospheric frames. Despite this known difficulty, the current distribution in the slice 'SL' during pre-to-post eruption (from top to bottom) phase, we note the degrading current concentrations, reflecting the indications of the field transition from non-potential sheared arcade to potential field. Owing to this fact, the energy estimations, as for the earlier event, also implies a similar result of lowering free-energy. 

In the top panel of Figure~\ref{Fig11}, we plot the horizontally integrated total magnetic energy and potential energy as a function of height, before (06:00UT on September 10 2014) and after (20:00UT on September 10 2014) the eruption. Because the extrapolation problem is a boundary dependent, the chosen times for this energy calculation are according to the time profile of the net vertical current (Figure~\ref{Fig3}, even $\alpha_{av}$) where a drastic decrease of net vertical current in both polarity regions is observed. Consistent with the current distribution, the free-energy dominantly lies within 40Mm, peaking at around 6 Mm. The free-energy curve after eruption is well below to that before eruption as a fact that the field is relaxed and less non-potential. During this static field evolution at the background of the observed dynamic eruption, the energy release is estimated as $1.05\times10^{32}$erg, which is sufficient to power GOES class X1.6 magnitude flare.            

\section{Summary and Discussion}
\label{sec4}
In this work, we investigated the relation of sunspot rotation with the major eruptions occurred in the vicinity of it. Vortex like motions are modelled to be potential triggers of eruptions in the ARs \citep{amari1996, torok2003, torok2013} by progressively twisting the line-tied foot points. Particularly, they involve in the formation or development of twisted flux ropes, and sigmoids by injecting twist and energy in the AR magnetic structure (e.g., \citealt{ruang2014}). We found that the location of the sunspot rotation had the magnetic roots of the erupting sigmoid (associated to two CME eruptions) that existed along the PIL between the sunspot and the surrounding opposite polarity in the AR 12158 \citep{chengx2015}. Like the earlier reports \citep{vemareddy2012b}, the correspondence of sunspot rotation motion is obviously reflected in the many non-potential parameters (Figure~\ref{Fig3}) during the evolution of the AR. Unlike the earlier cases, this AR is in the post-emerged phase with decreasing flux content, which reveals a direct role of observed sunspot rotation, a purely surface phenomenon of sub-photospheric origin, in the two major sigmoid eruptions. 

Since the driven motion by sunspot rotation is slow (of the order of 1km/s), the evolution of the magnetic structure is said to be quasi-static and therefore the evolution is approximated by a time series of force-free equilibria (e.g., \citealt{xudong2013, vemareddy2014a}). Under this scenario, utilizing HMI 12 minute cadence vector magnetic field observations, the AR magnetic structure is reconstructed by NLFF model, which reproduces the global structure in resemblance with the coronal EUV plasma structure. The modeled magnetic structure around the rotating sunspot appears as a fan-like sheared arcade, manifesting the observed sigmoid. Acknowledging the difficulty in working with the noisy observations \citep{hoeksema2014}, and also tracing the same structure in all time snapshots, the modeled field indicates signatures of accumulating strong coronal current concentrations and a building sigmoid at different time instances. 

While sunspot being observed to be rotating, a moderate CME eruption occurred at 23:00UT on September 8, 2014. During the onset of eruption, the AIA multi-thermal observations conspicuously present a continuous trace of hot flux rope embedded in the middle of ambient less hot sheared structure \citep{zhangj2012, chengx2015, vemareddy2014b}. The eruption is a partial one, where flare reconnection takes place slowly and accordingly a low speed CME associated with a long duration M4.6 flare is observed. Consistent with the photospheric measurement of net vertical current during pre-to-post eruption, an increased coronal current concentration is being observed across the sigmoid as a fact of twisting by sunspot rotation. As per these notices, the estimated free energy during the eruption is small, which we believe to be averaging effect because energy releases locally during field reconfiguration.

A second CME eruption launched at 17:30UT on September 10, 2014 in the AR. The CME is a halo heading toward Earth at a high speed (1014km/s) and follows an X1.6 flare. Appearance of continuous flux rope is evident amid the sigmoid during the onset of the eruption. As per net vertical current during this eruption event, the coronal current concentrations degraded across the sigmoid and the free energy estimation indicates a release of $1.44\times10^{31}$erg, which is sufficient for an X-class flare. These analysis results suggest that the magnetic connections of the sigmoid are driven by slow motion of sunspot rotation, which developed to a highly twisted flux rope structure in a dynamical scenario. An exceeding critical twist in the flux rope explains the loss of equilibrium triggering the onset of the observed eruptions. Although the NLFFF extrapolation worked best in reproducing highly twisted structure around the sunspot, given the limitations of both the observations and the model, a realization of a clear flux rope structure which is dynamic in nature, seem to be difficult task. Data driven MHD based models (e.g., \citealt{wust2006,jiangc2012, jiangc2014}) would help in better explaining the observed features in the eruptions under the influence of sunspot rotation, which would be the subject of our future investigations. 

\acknowledgements SDO is a mission of NASA's Living With a Star Program. This work used the DAVE4VM code, written and developed by P. W. Schuck at the Naval Research Laboratory. 3D rendering is due to VAPOR (\url{www.vapor.ucar.edu}) software. The authors sincerely thank Dr. T. Wiegelmann for providing NLFFF code. We acknowledge an extensive usage of the multi-node, multi-processor high performance computing facility at IIA. P.V. is supported by an INSPIRE grant of AORC scheme under the Department of Science and Technology. X.C. is supported by NSFC under grants 11303016. We thank the referee for constructive comments and suggestions that helped improving the manuscript greatly.

\bibliographystyle{apj}

\begin{thebibliography}{58}
\expandafter\ifx\csname natexlab\endcsname\relax\def\natexlab#1{#1}\fi

\bibitem[{{Amari} {et~al.}(2010){Amari}, {Aly}, {Mikic}, \&
  {Linker}}]{amari2010}
{Amari}, T., {Aly}, J.-J., {Mikic}, Z., \& {Linker}, J. 2010, \apjl, 717, L26

\bibitem[{{Amari} {et~al.}(1996){Amari}, {Luciani}, {Aly}, \&
  {Tagger}}]{amari1996}
{Amari}, T., {Luciani}, J.~F., {Aly}, J.~J., \& {Tagger}, M. 1996, \apjl, 466,
  L39

\bibitem[{{Antiochos} {et~al.}(1999){Antiochos}, {DeVore}, \&
  {Klimchuk}}]{antiochos1999}
{Antiochos}, S.~K., {DeVore}, C.~R., \& {Klimchuk}, J.~A. 1999, \apj, 510, 485

\bibitem[{{Aulanier} {et~al.}(2010){Aulanier}, {T{\"o}r{\"o}k}, {D{\'e}moulin},
  \& {DeLuca}}]{aulanier2010}
{Aulanier}, G., {T{\"o}r{\"o}k}, T., {D{\'e}moulin}, P., \& {DeLuca}, E.~E.
  2010, \apj, 708, 314

\bibitem[{{Barnes} \& {Sturrock}(1972)}]{barnes1972}
{Barnes}, C.~W., \& {Sturrock}, P.~A. 1972, \apj, 174, 659

\bibitem[{{Berger} \& {Field}(1984)}]{berger1984}
{Berger}, M.~A., \& {Field}, G.~B. 1984, Journal of Fluid Mechanics, 147, 133

\bibitem[{{Bhatnagar}(1967)}]{bhatnagar1967}
{Bhatnagar}, A. 1967, Kodaikanal Observ. Bull

\bibitem[{{Bobra} {et~al.}(2014){Bobra}, {Sun}, {Hoeksema}, \& {et
  al}}]{bobra2014}
{Bobra}, M.~G., {Sun}, X., {Hoeksema}, J.~T., \& {et al}. 2014, \solphys, 289,
  3549

\bibitem[{{Brown} {et~al.}(2003){Brown}, {Nightingale}, {Alexander}, \& {et
  al}}]{brown2003}
{Brown}, D.~S., {Nightingale}, R.~W., {Alexander}, D., \& {et al}. 2003,
  \solphys, 216, 79

\bibitem[{{Cheng} {et~al.}(2015){Cheng}, {Ding}, \& {Fang}}]{chengx2015}
{Cheng}, X., {Ding}, M.~D., \& {Fang}, C. 2015, \apj, 804, 82

\bibitem[{{Cheng} {et~al.}(2013){Cheng}, {Zhang}, {Ding}, \& {et
  al}}]{chengx2013}
{Cheng}, X., {Zhang}, J., {Ding}, M.~D., \& {et al}. 2013, \apjl, 769, L25

\bibitem[{{Evershed}(1910)}]{evershed1910}
{Evershed}, J. 1910, \mnras, 70, 217

\bibitem[{{Galsgaard} \& {Nordlund}(1997)}]{galsgaard1997}
{Galsgaard}, K., \& {Nordlund}, {\AA}. 1997, \jgr, 102, 219

\bibitem[{{Gerrard} {et~al.}(2002){Gerrard}, {Arber}, \& {Hood}}]{garrard2002}
{Gerrard}, C.~L., {Arber}, T.~D., \& {Hood}, A.~W. 2002, \aap, 387, 687

\bibitem[{{Gibson} {et~al.}(2006){Gibson}, {Fan}, {T{\"o}r{\"o}k}, \&
  {Kliem}}]{gibson2006}
{Gibson}, S.~E., {Fan}, Y., {T{\"o}r{\"o}k}, T., \& {Kliem}, B. 2006, \ssr,
  124, 131

\bibitem[{{Gilbert} {et~al.}(2007){Gilbert}, {Alexander}, \&
  {Liu}}]{gilbert2007}
{Gilbert}, H.~R., {Alexander}, D., \& {Liu}, R. 2007, \solphys, 245, 287

\bibitem[{{Hagino} \& {Sakurai}(2004)}]{hagino2004}
{Hagino}, M., \& {Sakurai}, T. 2004, \pasj, 56, 831

\bibitem[{{Hiremath} \& {Suryanarayana}(2003)}]{hiremath2003}
{Hiremath}, K.~M., \& {Suryanarayana}, G.~S. 2003, \aap, 411, L497

\bibitem[{{Hoeksema} {et~al.}(2014){Hoeksema}, {Liu}, {Hayashi}, {Sun}, \& {et
  al}}]{hoeksema2014}
{Hoeksema}, J.~T., {Liu}, Y., {Hayashi}, K., {Sun}, X., \& {et al}. 2014,
  \solphys, 289, 3483

\bibitem[{{Jiang} {et~al.}(2012{\natexlab{a}}){Jiang}, {Feng}, {Wu}, \&
  {Hu}}]{jiangc2012}
{Jiang}, C., {Feng}, X., {Wu}, S.~T., \& {Hu}, Q. 2012{\natexlab{a}}, \apj,
  759, 85

\bibitem[{{Jiang} {et~al.}(2014){Jiang}, {Wu}, {Feng}, \& {Hu}}]{jiangc2014}
{Jiang}, C., {Wu}, S.~T., {Feng}, X., \& {Hu}, Q. 2014, \apj, 780, 55

\bibitem[{{Jiang} {et~al.}(2012{\natexlab{b}}){Jiang}, {Zheng}, {Yang}, {Hong},
  {Yi}, \& {Yang}}]{jiangy2012}
{Jiang}, Y., {Zheng}, R., {Yang}, J., {Hong}, J., {Yi}, B., \& {Yang}, D.
  2012{\natexlab{b}}, \apj, 744, 50

\bibitem[{{Kazachenko} {et~al.}(2009){Kazachenko}, {Canfield}, {Longcope}, \&
  {et al}}]{kazachenko2009}
{Kazachenko}, M.~D., {Canfield}, R.~C., {Longcope}, D.~W., \& {et al}. 2009,
  \apj, 704, 1146

\bibitem[{{Kusano} {et~al.}(2002){Kusano}, {Maeshiro}, {Yokoyama}, \&
  {Sakurai}}]{kusano2002}
{Kusano}, K., {Maeshiro}, T., {Yokoyama}, T., \& {Sakurai}, T. 2002, \apj, 577,
  501
	
\bibitem[{Leamon} {et~al.}(2003)]{leamon2003} 
{Leamon}, R.~J., {Canfield}, R.~C., {Blehm}, Z., \& {Pevtsov}, A.~A.\ 2003, \apjl, 596, L255 

\bibitem[{{Lemen} {et~al.}(2012){Lemen}, {Title}, {Akin}, {Boerner}, \& {et
  al}}]{lemen2012}
{Lemen}, J.~R., {Title}, A.~M., {Akin}, D.~J., {Boerner}, P.~F., \& {et al}.
  2012, \solphys, 275, 17

\bibitem[{{Liu} \& {Schuck}(2012)}]{liuy2012}
{Liu}, Y., \& {Schuck}, P.~W. 2012, \apj, 761, 105

\bibitem[{{Mackay} {et~al.}(2011){Mackay}, {Green}, \& {van
  Ballegooijen}}]{mackay2011}
{Mackay}, D.~H., {Green}, L.~M., \& {van Ballegooijen}, A. 2011, \apj, 729, 97

\bibitem[{{McIntosh}(1981)}]{mcintosh1981}
{McIntosh}, P.~S. 1981, in The Physics of Sunspots, ed. L.~E. {Cram} \& J.~H.
  {Thomas}, 7--54

\bibitem[{{Mikic} {et~al.}(1990){Mikic}, {Schnack}, \& {van Hoven}}]{mikic1990}
{Mikic}, Z., {Schnack}, D.~D., \& {van Hoven}, G. 1990, \apj, 361, 690

\bibitem[{{Nindos} {et~al.}(2012){Nindos}, {Patsourakos}, \&
  {Wiegelmann}}]{nindos2012}
{Nindos}, A., {Patsourakos}, S., \& {Wiegelmann}, T. 2012, \apjl, 748, L6

\bibitem[{{Ruan} {et~al.}(2014){Ruan}, {Chen}, {Wang}, {Zhang}, {Li}, {Jing},
  {Su}, {Li}, {Xu}, {Du}, \& {Wang}}]{ruang2014}
{Ruan}, G., {Chen}, Y., {Wang}, S., {Zhang}, H., {Li}, G., {Jing}, J., {Su},
  J., {Li}, X., {Xu}, H., {Du}, G., \& {Wang}, H. 2014, \apj, 784, 165

\bibitem[{{Savcheva} \& {van Ballegooijen}(2009)}]{savcheva2009}
{Savcheva}, A., \& {van Ballegooijen}, A. 2009, \apj, 703, 1766

\bibitem[{{Schou} {et~al.}(2012){Schou}, {Scherrer}, {Bush}, {Wachter}, \& {et
  al}}]{schou2012}
{Schou}, J., {Scherrer}, P.~H., {Bush}, R.~I., {Wachter}, R., \& {et al}. 2012,
  \solphys, 275, 229

\bibitem[{{Schuck}(2008)}]{schuck2008}
{Schuck}, P.~W. 2008, \apj, 683, 1134

\bibitem[{{Stenflo}(1969)}]{stenflo1969}
{Stenflo}, J.~O. 1969, \solphys, 8, 115

\bibitem[{{Sun} {et~al.}(2013){Sun}, {Hoeksema}, {Liu}, {Aulanier}, {Su},
  {Hannah}, \& {Hock}}]{xudong2013}
{Sun}, X., {Hoeksema}, J.~T., {Liu}, Y., {Aulanier}, G., {Su}, Y., {Hannah},
  I.~G., \& {Hock}, R.~A. 2013, \apj, 778, 139

\bibitem[{{Sun} {et~al.}(2012){Sun}, {Hoeksema}, {Liu}, {Wiegelmann},
  {Hayashi}, {Chen}, \& {Thalmann}}]{xudong2012}
{Sun}, X., {Hoeksema}, J.~T., {Liu}, Y., {Wiegelmann}, T., {Hayashi}, K.,
  {Chen}, Q., \& {Thalmann}, J. 2012, \apj, 748, 77

\bibitem[{{Suryanarayana}(2010)}]{suryanarayana2010}
{Suryanarayana}, G.~S. 2010, New Astronomy, 15, 313

\bibitem[{{Tian} \& {Alexander}(2006)}]{tian2006}
{Tian}, L., \& {Alexander}, D. 2006, \solphys, 233, 29

\bibitem[{{Tian} {et~al.}(2008){Tian}, {Alexander}, \&
  {Nightingale}}]{tian2008}
{Tian}, L., {Alexander}, D., \& {Nightingale}, R. 2008, \apj, 684, 747

\bibitem[{{Titov} {et~al.}(2002){Titov}, {Hornig}, \&
  {D{\'e}moulin}}]{titov2002}
{Titov}, V.~S., {Hornig}, G., \& {D{\'e}moulin}, P. 2002, Journal of
  Geophysical Research (Space Physics), 107, 1164

\bibitem[{{Tokman} \& {Bellan}(2002)}]{tokman2002}
{Tokman}, M., \& {Bellan}, P.~M. 2002, \apj, 567, 1202

\bibitem[{{T{\"o}r{\"o}k} \& {Kliem}(2003)}]{torok2003}
{T{\"o}r{\"o}k}, T., \& {Kliem}, B. 2003, \aap, 406, 1043

\bibitem[{{T{\"o}r{\"o}k} \& {Kliem}(2005)}]{torok2005}
---. 2005, \apjl, 630, L97

\bibitem[{{T{\"o}r{\"o}k} {et~al.}(2013){T{\"o}r{\"o}k}, {Temmer}, {Valori},
  {Veronig}, {van Driel-Gesztelyi}, \& {Vr{\v s}nak}}]{torok2013}
{T{\"o}r{\"o}k}, T., {Temmer}, M., {Valori}, G., {Veronig}, A.~M., {van
  Driel-Gesztelyi}, L., \& {Vr{\v s}nak}, B. 2013, \solphys, 286, 453

\bibitem[{{Vemareddy}(2015)}]{vemareddy2015b}
{Vemareddy}, P. 2015, \apj, 806, 245

\bibitem[{{Vemareddy} {et~al.}(2012{\natexlab{a}}){Vemareddy}, {Ambastha}, \&
  {Maurya}}]{vemareddy2012b}
{Vemareddy}, P., {Ambastha}, A., \& {Maurya}, R.~A. 2012{\natexlab{a}}, \apj,
  761, 60

\bibitem[{{Vemareddy} {et~al.}(2012{\natexlab{b}}){Vemareddy}, {Maurya}, \&
  {Ambastha}}]{vemareddy2012a}
{Vemareddy}, P., {Maurya}, R.~A., \& {Ambastha}, A. 2012{\natexlab{b}},
  \solphys, 277, 337

\bibitem[{{Vemareddy} {et~al.}(2015){Vemareddy}, {Venkatakrishnan}, \&
  {Karthikreddy}}]{vemareddy2015a}
{Vemareddy}, P., {Venkatakrishnan}, P., \& {Karthikreddy}, S. 2015, Research in
  Astronomy and Astrophysics, 15, 1547

\bibitem[{{Vemareddy} \& {Wiegelmann}(2014)}]{vemareddy2014a}
{Vemareddy}, P., \& {Wiegelmann}, T. 2014, \apj, 792, 40

\bibitem[{{Vemareddy} \& {Zhang}(2014)}]{vemareddy2014b}
{Vemareddy}, P., \& {Zhang}, J. 2014, \apj, 797, 80

\bibitem[Wiegelmann(2004)]{wiegelmann2004} 
{Wiegelmann}, T.\ 2004, \solphys, 219, 87 

\bibitem[{Wiegelmann} \& {Inhester}(2010)]{wiegelmann2010} 
{Wiegelmann}, T., \& {Inhester}, B.\ 2010, \aap, 516, A107 

\bibitem[{{Wiegelmann} {et~al.}(2006){Wiegelmann}, {Inhester}, \&
  {Sakurai}}]{wiegelmann2006}
{Wiegelmann}, T., {Inhester}, B., \& {Sakurai}, T. 2006, \solphys, 233, 215

\bibitem[{{Wu} {et~al.}(2006){Wu}, {Wang}, {Liu}, \& {Hoeksema}}]{wust2006}
{Wu}, S.~T., {Wang}, A.~H., {Liu}, Y., \& {Hoeksema}, J.~T. 2006, \apj, 652,
  800

\bibitem[{{Yan} {et~al.}(2008){Yan}, {Qu}, \& {Kong}}]{yanxl2008}
{Yan}, X.-L., {Qu}, Z.-Q., \& {Kong}, D.-F. 2008, \mnras, 391, 1887

\bibitem[{{Zhang} {et~al.}(2012){Zhang}, {Cheng}, \& {Ding}}]{zhangj2012}
{Zhang}, J., {Cheng}, X., \& {Ding}, M.-D. 2012, Nature Communications, 3

\bibitem[{{Zhang} {et~al.}(2007){Zhang}, {Li}, \& {Song}}]{zhangj2007}
{Zhang}, J., {Li}, L., \& {Song}, Q. 2007, \apjl, 662, L35

\bibitem[{{Zhang} {et~al.}(2008){Zhang}, {Liu}, \& {Zhang}}]{zhangy2008}
{Zhang}, Y., {Liu}, J., \& {Zhang}, H. 2008, \solphys, 247, 39

\bibitem[{{Zhao} {et~al.}(2016){Zhao}, {Gilchrist}, {Aulanier}, {Schmieder},
  {Pariat}, \& {Li}}]{zhaoj2016}
{Zhao}, J., {Gilchrist}, S.~A., {Aulanier}, G., {Schmieder}, B., {Pariat}, E.,
  \& {Li}, H. 2016, ArXiv e-prints

\end{thebibliography}


\end{document}